\documentclass[structabstract]{aa}

\usepackage{txfonts}
\usepackage{graphicx}
\usepackage{natbib}
\bibpunct{(}{)}{;}{a}{}{,} % to follow the A&A style

\newcommand{\ME}{M$_{\oplus}$}

\newcommand{\MJ}{M$_\mathrm{J}$}
\newcommand{\RJ}{R$_\mathrm{J}$}

\begin{document}
 
\title{Constraining the interior of extrasolar giant planets with 
  the tidal Love number $k_2$ using the example of HAT-P-13b}

\author{U. Kramm\inst{\ref{UniHRO}}\thanks{\email{ulrike.kramm2@uni-rostock.de}}
\and N. Nettelmann\inst{\ref{UniHRO}} \and J. J. Fortney\inst{\ref{UCSC}} \and
R. Neuh\"auser\inst{\ref{Jena}} \and R. Redmer\inst{\ref{UniHRO}}}

\institute{Institute of Physics, University of Rostock, D-18051 Rostock
\label{UniHRO}
\and
Dept. of Astronomy and Astrophysics, University of California, Santa
Cruz, CA 95064\label{UCSC}
\and
Astrophysical Institute and University-Observatory, Schillerg\"asschen 2-3,
D-07745 Jena \label{Jena}}

\date{Received ... / Accepted ...}

\abstract
{Transit and radial velocity observations continuously discover an increasing
number of exoplanets. However, when it comes to the composition of the observed
planets the data are compatible with several interior structure
models. Thus, a planetary parameter sensitive to the planet's density
distribution could help constrain this large number of
possible models even further.}
{We aim to investigate to what extent an exoplanet's interior can
be constrained in terms of core mass and envelope metallicity by taking
the
tidal Love number $k_2$ into account as an additional,
possibly observable parameter.}
{Because it is the only planet with an observationally determined
$k_2$, we
constructed interior models for the Hot Jupiter
exoplanet HAT-P-13b by solving
the equations of hydrostatic equilibrium and mass conservation for different
boundary conditions. In particular, we varied the surface temperature
and the
outer temperature profile, as well as the envelope metallicity within the widest
possible parameter range. We also considered atmospheric conditions
that are
consistent with nongray atmosphere models. For all these models we
calculated the
Love number $k_2$ and compared it to the allowed range of $k_2$ values
that could
be obtained from eccentricity measurements of HAT-P-13b.}
{We use the example of HAT-P-13b to show the general relationships
between the quantities temperature, envelope metallicity, core mass, and Love
number of a planet. For any given $k_2$ value a maximum possible core
mass can be
determined. For HAT-P-13b we find $M_\mathrm{core}<27$\,\ME, based on the latest
eccentricity measurement. We favor models that are consistent with our model
atmosphere, which gives us the
temperature of the isothermal region as $\sim2100$\,K.
With this external
boundary condition and our new $k_2$-interval we are able to constrain
both the envelope and bulk
metallicity of HAT-P-13b to 1 -- 11 times stellar metallicity
and the
extension of the isothermal layer in the planet's atmosphere to 3 -- 44\,bar.
Assuming equilibrium tidal theory, we find lower limits on the tidal $Q$
consistent with $10^3 - 10^5$.}
{Our analysis shows that the tidal Love number $k_2$ is a very useful
parameter for studying the interior of exoplanets. It allows one to
place limits on the core mass
and estimate the metallicity of a planet's envelope.}


\titlerunning{Constraining the interior of exoplanets with the tidal Love number
$k_2$}
\authorrunning{U. Kramm et al.}

\maketitle

\section{Introduction}
Today more than 180 transiting exoplanets have been discovered. Within
the
exoplanet family planets that are found to transit their host star are
especially important. The knowledge of mass \emph{and} radius enables us to
infer the density and bulk composition of a planet. Further, transit
and secondary eclipse observations reveal information about an exoplanet's
atmosphere.

Despite all the information given, we are still far away from knowing the
composition of the deep interior of exoplanets. Characteristics like
the existence and mass of a potential core or the amount and
distribution of heavy
elements are quite ambiguous. Yet, this information is highly desired
in order to understand planet formation.

For the solar system planets, the ambiguity of interior models can be reduced
by taking into account information from the gravity field, which is quantified
by the gravitational moments $J_2$, $J_4$, and $J_6$. In our solar system these
parameters are measureable and improved methods continuously increase their
accuracy. However, for an extrasolar planet gravitational moments cannot be
determined. Hence, we need a similar parameter that is accessible and will also
provide us with information about the interior density distribution of the
planet. This can be accomplished with the tidal Love number $k_2$
\citep{Love11,Gavrilovetal75,GavrilovZharkov77,ZharkovTrubitsyn78}. As it is
equivalent to $J_2$ for the solar system planets \citep{Hubbard84}, it is
promising that $k_2$
will help us to further constrain the interior models of exoplanets. Recent
studies have shown that $k_2$ is itself in general a degenerate quantity
if
considered in a three-layer model \citep{Krammetal11}.

In this study, we focus on the transiting Hot Jupiter HAT-P-13b
\citep{Bakosetal09}. This planet is of great interest because it is the
only planet so far which has an
observationally determined value for the Love number $k_2$
\citep{Batyginetal09}. Hence, it gives us the chance to investigate how much
information about the interior structure can be inferred from an actually
measured $k_2$ and serves as a case study in this work. We especially
evaluate the impact of the new eccentricity measurement for HAT-P-13b by
\citet{Winnetal10} as it gives new information about allowed $k_2$ values.

In \S\ref{sec:Methods} we describe our modeling procedure, the equation of
state (EOS) used, and our calculation of the Love number.
The known observational constraints of the examined planet HAT-P-13b are
given in \S\ref{sec:inputmodelassumptions} where we also discuss the effect of
different
eccentricity measurements, and describe the setup of our interior models.
The results are explained in \S\ref{sec:Results}, and in \S\ref{sec:Discussion}
we compare with previous results. The applicability of the
underlying theory is discussed in the Appendix. A summary is given in
\S\ref{sec:Summary}.

\section{Methods} \label{sec:Methods}

\subsection{Modeling procedure}

We construct planetary interior models by integrating the equation of
hydrostatic equilibrium
\begin{equation}
 \frac{\mathrm{d}P}{\mathrm{d}r}=-\frac{Gm(r)}{r^2}\rho(r)
\label{eq:hydrostaticequilibrium}
\end{equation}
and the equation of mass conservation
\begin{equation}
 M_\mathrm{p}=\int_0^{R_\mathrm{p}}4\pi r^2 \rho(r)\,\mathrm{d}r
\end{equation}
inward, where $P$ is the pressure, $r$ the radial coordinate, $\rho(r)$ the mass
density at $r$, $m(r)$ the mass inside the radius $r$, and $M_\mathrm{p}$ and
$R_\mathrm{p}$ the total mass and radius of the planet, respectively. We
define the surface of the planet $R_\mathrm{p}$ at the 1-bar pressure
level
$P_1=1\,\mathrm{bar}$. For Hot Jupiters the measured transit radius may be at
lower pressures of about $\sim 10$\,mbar 
\citep{Fortneyetal03}. For HAT-P-13b, however, the very outermost layer from
1\,bar to 10\,mbar does not contribute significantly to the radius (within the
error bars for the transit radius measurement). Hence, we start
our calculation of the interior at the 1-bar pressure level as
previously done for solar system planets \citep{Guillot99}.
The equations above require the following outer boundary conditions: the radius
inside of $P_1$ equals the planet radius, so $r(P_1)=R_\mathrm{p}$, and the mass
inside of $R_\mathrm{p}$ is equal to the planet's mass, so
$m(R_\mathrm{p})=M_\mathrm{p}$. Furthermore, a planet has to
obey the inner boundary condition $m(r=0)=0$. To fulfill this condition, we
assume that a core exists and choose the core mass ($M_\mathrm{core}\geq0$) such
that total mass conservation is ensured.
Equation~(\ref{eq:hydrostaticequilibrium}) needs a $P$-$\rho$-relation
(equation of state). For details about the EOS used, see \S~\ref{subsec:EOS}.
To obtain $\rho_1=\rho(P_1,T_1)$ from the EOS we need the surface temperature
$T_1$ which we consider as a variable parameter.

\subsection{EOS} \label{subsec:EOS}

For the interior models in this work we assume a composition of hydrogen,
helium, metals and rocks.
The rocks are confined to a core and we apply the pressure-density
relation by \citet{HubbardMarley89} which approximates a mixture of 38\%
SiO$_2$, 25\% MgO, 25\% FeS, and 12\% FeO.
For the envelope we suppose a mixture of H, He, and metals.
For H and He we use the interpolated SCvH-i EOS \citep{Saumonetal95}.
To represent elements heavier than He (metals) we scale the density of the
He-EOS by a factor of four.

\subsection{Love number calculation}

For the interior models generated in this work we calculate the Love
number $k_2$ for a hydrostatic planet. This planetary property quantifies the
deformation of the
quadrupolic gravity field of the planet in response to an external massive body.
In the cases of exoplanetary systems of interest, the parent star of mass $M_*$
causes a tide-raising potential \citep{ZharkovTrubitsyn78}
\begin{equation}
 W(s)=\sum^\infty_{n=2}W_n=\frac{GM_*}{a}\sum^\infty_{n=2}\left(\frac{s}{a}
\right)^nP_n(\cos\theta')\quad,
\end{equation}
where $a$ is the radius of the planet's orbit, $s$ the radial coordinate of a
point inside the planet, $\theta'$ the angle between a planetary mass element
at $s$ and $M_*$ at $a$, and $P_n$ are Legendre polynomials. The response of
the planet's potential at the surface is given by
\begin{equation}
 V_n^\mathrm{ind}(R_\mathrm{p})=k_nW_n(R_\mathrm{p})\quad,
\end{equation}
where $R_\mathrm{p}$ is the radius of the planet and $k_n$ are its tidal Love
numbers. Like the gravitational moments $J_{2n}$ they are determined by the
planet's internal density distribution. For a more detailed description of the
method used to calculate Love numbers and an analysis about some general
dependences of $k_2$ on the density distribution see \citet{Krammetal11} and
references therein.

For the solar system giant planets the lowest order gravitational moments
$J_2$ and $J_4$ have proven to be extremely valuable constraints for
planet modeling \citep[see e.g.][]{Guillot99,SaumonGuillot04,Nettelmann11}. For
extrasolar planets on the other hand, constraints derived from transit and
radial
velocity observations are often limited to the mass, radius and effective
temperature of the planet. In a few cases information about the atmosphere can
be obtained from spectroscopy \citep[see e.g][]{Charbonneauetal02, Pontetal08},
which at the current level helps to validate atmosphere models
\citep{Fortneyetal10}.

It can be shown that to first order in the expansion of the planet's potential,
the Love number $k_2$ is proportional to $J_2$ \citep[see e.g.][]{Hubbard84},
indicating that a measurement of $k_2$ provides us with equivalent information
like $J_2$ for the solar system giants.
In this way, knowledge about the Love number $k_2$ can give us some insight into
the planet's interior by constituting an additional constraint if observed. In
fact, an estimation of the tidal Love number $k_2$ based on observations was
made for the Hot Jupiter HAT-P-13b, as it is part of a system in a \emph{tidal
fixed point} (see \S\ref{subsec:HATP13b-obsconst}). The consequences
of this $k_2$-determination for the interior of HAT-P-13b are discussed in
\S\ref{sec:Results}. Another possibility to determine $k_2$
observationally is to detect the tidally induced apsidal precession that is
expected to be seen in the transit light curves of Hot Jupiters
\citep{RagozzineWolf09}.

\section{Input data and model assumptions} \label{sec:inputmodelassumptions}

\subsection{Observational constraints} \label{subsec:HATP13b-obsconst}

%discovery Bakosetal09
Discovered in 2009 by \citet{Bakosetal09}, the HAT-P-13 system is the first
exoplanetary system that contains a transiting planet that is accompanied by a
well-characterized longer-period companion. The planets orbit a metal
rich ($\mathrm{[Fe/H]}=+0.41\pm0.08\,\mathrm{dex}$) G4 star with
$L_*=2.22\pm0.31\,L_\odot$. The inner planet HAT-P-13b is a transiting Hot
Jupiter which
passes in front of its star every $P_b=2.916260\pm0.000010$ days, correspondig
to an orbital distance of $a_b=0.0427^{+0.0006}_{-0.0012}$\,AU. The orbit is
nearly circular with $e_b=0.021\pm0.009$\footnote{The value of the
eccentricity is actually a matter of debate and influences the results
significantly (see \S\ref{subsec:e}).}. The radius and mass of HAT-P-13b were
determined with transit and radial velocity follow-up observations to be
$M_\mathrm{p}=0.853^{+0.029}_{-0.046}$\,\MJ, and $R_\mathrm{p}=
1.281\pm0.079$\,\RJ, respectiveley.
The companion HAT-P-13c is on a more distant orbit at
$a_c=1.188^{+0.018}_{-0.033}$\,AU with a period of $P_c=428.5\pm3.0$ days. As it
has not been observed in transit \citep{Szaboetal10} only the minimum mass
$M\sin i=15.2\pm1.0$\,\MJ~is known. An important feature of the system is that
the c planet\footnote{With its high mass the companion may instead be
classified as a
brown dwarf. However, as the definition of planets and brown dwarfs is
rather unclear
\citep{Baraffeetal08}, we call it a planet here.} is on a highly eccentric
orbit with $e_c=0.691\pm0.018$. For
more details about the observed parameters of the system see
\citet{Bakosetal09}.

%tidal evolution Mardling
The high eccentricity of planet c (compared to the inner short-period planet) is
one prerequisite for the theory of \citet{Mardling07} on the tidal evolution
of a two-planet system. A further requirement is that the system must be
co-planar. Then, according to Mardling's theory, the system evolves in the
following phases:
(1) The angle $\eta$ between the apsidal lines of the two planets circulates and
the eccentricities slowly oscillate at constant amplitude accompanied by a
decrease of the mean value of the inner planet's eccentricity. This occurs on
the circularization time scale $\tau_\mathrm{circ}$. For HAT-P-13b
$\tau_\mathrm{circ} \sim 0.05$\,Gyr assuming $Q_\mathrm{p}\sim10^5$ and
$k_2\sim 0.3$.
(2) $\eta$ librates around a fixed value $\eta = 0, \pi$. The
eccentricities slowly oscillate with declining amplitude but maintain the
mean value of the inner planet's eccentricity. This occurs on twice the
circularization time scale. At the end of this phase the system resides in the
\emph{tidal fixed point} characterized by aligned apsidal lines which then
precess with the same rate.
(3) The subsequent long term evolution  proceeds on a time scale several orders
of magnitude longer than the circularization time scale. So the tidal fixed
point is actually a \emph{quasi fixed point}. Long term tidal effects will
cause a slow non-oscillatory decline of both eccentricities to zero unless the
companion is not of very low mass and long period.

%k2 from that Batygin et al.
\citet{Bakosetal09} observed that the pericenter of
HAT-P-13b is aligned with that of the outer planet HAT-P-13c to within
$4^\circ\pm40^\circ$. If the apsidal lines are indeed aligned and if the system
is co-planar, it is in the tidal fixed point. \citet{Batyginetal09} (hereafter
referred to as BBL) showed that under this assumption one can estimate the tidal
Love number $k_2$ for the inner planet which in turn will give information about
the interior of planet b. They first generated models that satisfy
a given choice of the planetary mass, the radius, and an inferred planetary
effective
temperature $T_\mathrm{eff}=1649$\,K. For these models BBL then calculated
$k_2$ from the density distribution $\rho(r)$. Furthermore, they determined the
fixed point eccentricity $e_b$ based on the constraint that the orbits of both
planets have an identical precession rate when the system is in the tidal fixed
point configuration:
$\dot{\overline{\omega}}_{b\mathrm{total}}(e_b,k_2) = \dot{\overline{\omega}}_{
c\mathrm{secular} } $ , where $\dot{\overline{\omega}}_{b\mathrm{total}}$ is the
total precession of the inner planet incorporating the four most significant
contributions arising from secular evolution induced by planet-planet
interaction, the tidal and rotational bulges of the planet, and general
relativity, where $\dot{\overline{\omega}}_{c\mathrm{secular}}$ is the
precession
of the orbit of planet c due to the secular evolution \citep[see
also][]{Mardling07}. BBL found that the different $k_2$-values of the inner
planet result in significantly different eccentricities $e_b$ with larger $e_b$
implying a smaller $k_2$ and vice versa. They approximated this behavior with a
fourth-order polynomial $e_b(k_2)$. In summary, if a planetary
system fulfills the requirements for the tidal evolution theory by
\citet{Mardling07} and has evolved into the tidal fixed point, one can estimate
the Love number $k_2$ of the inner planet by measuring the orbital parameters
of the planets. It is important to note that the estimates for $k_2$
are only valid under these specific conditions, which may not
apply for
HAT-P-13 (see Appendix~\ref{sec:appfixedpoint} for a discussion of the
observational evidence of the requirements for the tidal fixed point theory).
Nevertheless, HAT-P-13b serves as
a good example for us here to investigate what conclusions about the interior of
an extrasolar giant planet can be drawn from an inferred $k_2$ value.

In the case of HAT-P-13b, BBL derived from the measured
$e_b=0.021\pm0.009$ an allowed $k_2$-interval of $0.116<k_2<0.425$ (hereafter
denoted as BBL-interval). They further concluded that the core mass of
HAT-P-13b is 0\,\ME\,$<M_c<$\,120\,\ME.

\subsection{The importance of the eccentricity measurement} \label{subsec:e}

\begin{figure}
 \resizebox{\hsize}{!}{\includegraphics{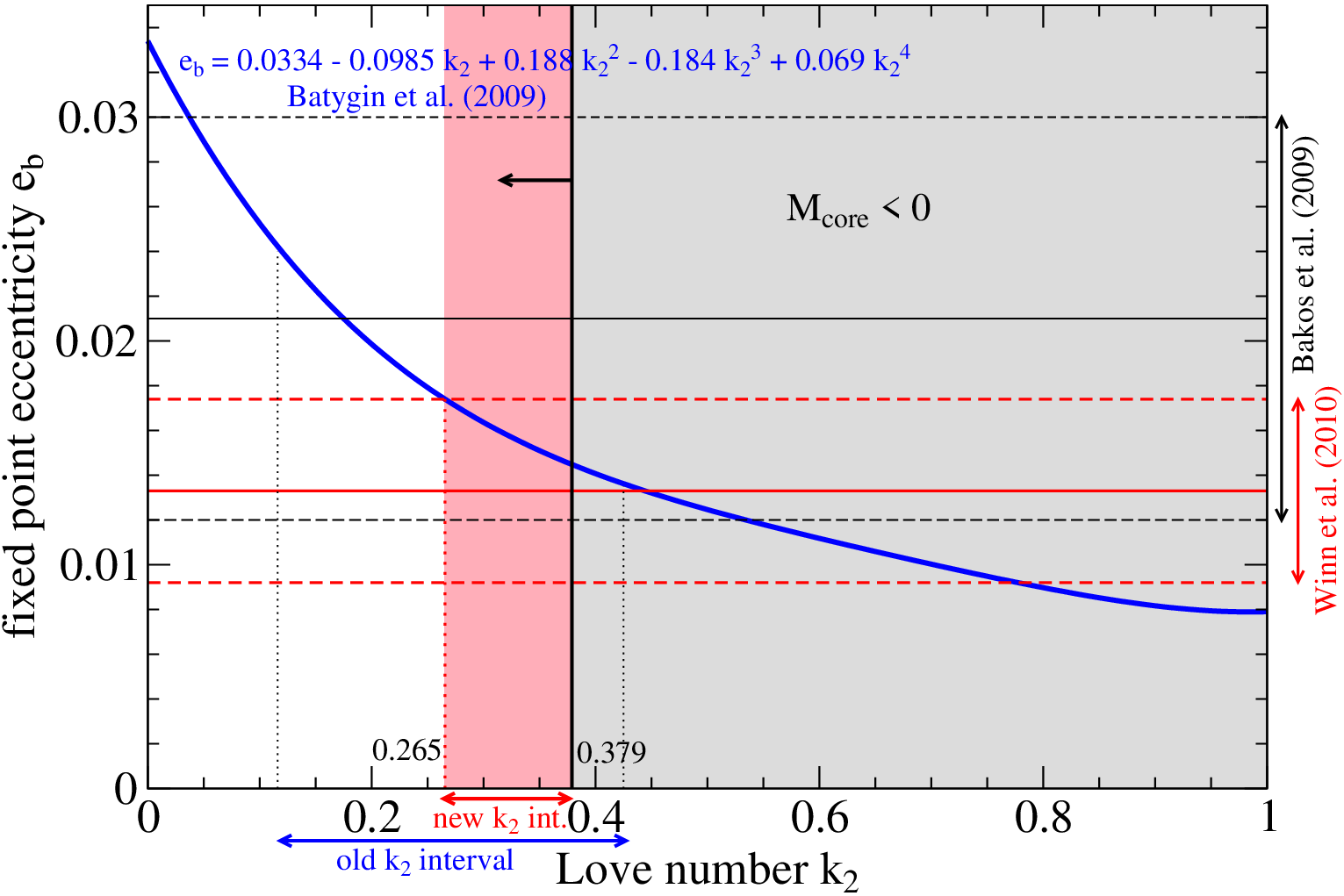}}
 \caption{Relation between the fixed point eccentricity $e_b$ and the Love
number $k_2$. The blue line shows the fourth order polynomial fit from
\citet{Batyginetal09}. Different eccentricity measurements from
\citet{Bakosetal09} and \citet{Winnetal10} are plotted in black and red,
respectively, with solid lines showing the mean value of the measured
eccentricity and dashed lines showing the error bars. The black dotted lines
indicate the $k_2$ interval inferred from \citet{Batyginetal09} based on the
eccentricity measurement from \citet{Bakosetal09}, in the figure also referred
to as \emph{old $k_2$ interval}. The gray shaded area marks a region of
$k_2$ values
not possible for models of HAT-P-13b. The vertical black solid line
($M_\mathrm{core}=0$) shows the
maximum possible $k_2$ value we found based on our interior modeling (see
\S\ref{sec:Results} for details). Combining the information of the
$e_b$-$k_2$-relation from \citet{Batyginetal09}, the new eccentricity
measurement from \citet{Winnetal10}, and our interior modeling, we find a
\emph{new $k_2$ interval} ranging from 0.265 -- 0.379 (rosy shaded).}
 \label{fig:e}
\end{figure}

The inner planet's eccentricity is a crucial observable in our analysis because
this quantity determines the allowed $k_2$ interval. In fact, the value of the
eccentricity is somewhat uncertain. While \citet{Bakosetal09} report
$e_b=0.021\pm0.009$, a later study of new radial velocity measurements by
\citet{Winnetal10} found an eccentricity of only half that value
($e_b=0.0133\pm0.0041$). We favor the eccentricity measurement from
\citet{Winnetal10}, because they include the data from \citet{Bakosetal09} and,
in addition, use 75 new RV data points which extend the timespan of the data set
by about 1 yr, and therefore allow a refinement of the orbital parameters. The
consequences for the inferred $k_2$ interval are
illustrated in Fig.~\ref{fig:e}.

Based on the eccentricity measurement from \citet{Bakosetal09}, BBL found an
allowed interval for $k_2$ of 0.116 -- 0.425. This result also includes
information from their interior modeling, which is why the limits of the
interval are not equal to the intersections of their fourth order polynomial
fit and the error bars of the eccentricity measurement. \citet{Winnetal10}
found a significantly smaller eccentricity. As the eccentricity gets
smaller, the resulting $k_2$
must become larger. As a consequence, the lower limit of the $k_2$
interval is
raised to 0.265. To find the upper limit, we make use of our interior models.
We find the most homogeneous model of HAT-P-13b at a maximum $k_2$ of 0.379. At
this point the core mass vanishes and there are no solutions beyond that $k_2$
value (for details see \S\ref{sec:Results}). This upper limit is also
insensitive to variations in mass and radius within the observational error bars
(see \S~\ref{subsec:MRuncertainties}).

By combining the information of the fourth order polynomial fit from BBL, the
new eccentricity measurement from \citet{Winnetal10}, and our interior models,
we derived a \emph{new $k_2$ interval} of $0.265 < k_2 < 0.379$. It can be
expected that this smaller region of allowed $k_2$-values will significantly
narrow
the allowed range of solutions for the interior of HAT-P-13b.

\subsection{Model assumptions} \label{subsec:Modelassumptions}

For our models we use the mean values for the mass and radius of HAT-P-13b
\citep{Bakosetal09}:
\[ M_\mathrm{p}=0.853\,\mathrm{M}_\mathrm{J}=271\,\mathrm{M}_\oplus\quad,\]
\[ R_\mathrm{p}=1.281\,\mathrm{R}_\mathrm{J}=14.35\,\mathrm{R}_\oplus\quad.\]
The uncertainties introduced by the error bars in mass and radius are
discussed in \S~\ref{subsec:MRuncertainties}.
We assume a two-layer structure, consisting of a rocky core and one homogeneous
envelope of hydrogen, helium and metals. It is not necessary to consider
three-layer models in this work because they are within the range of solutions
confined by the two-layer models (see \S\ref{sec:Results} for a
discussion).
We set the helium abundance in the envelope to the solar value $Y=0.27$
\citep{Bahcalletal95}. The amount of heavy elements $Z$ was varied from
$Z=0$, which gives a model with the biggest possible core when the other
parameters are fixed, to a maximum value where the core vanishes.

Due to the proximity to its star, the temperature profile of HAT-P-13b is
likely to be strongly influenced by the radiation of the star. The various
processes in an exoplanet's atmosphere influencing its temperature are very
complex and beyond the scope of this work. As we aim to study the effects of
input parameters important for planet modeling on the structure and Love
number of a planet, we vary the envelope temperature within the widest possible
range, even though extremely cold or hot models may be unrealistic. For
HAT-P-13b we change the 1\,kbar temperatures from 3800 to 9000\,K.
$T_{1\,\mathrm{kbar}}=3800$\,K gives the most homogeneous planet with zero core
mass and models with $T_{1\,\mathrm{kbar}}>9000$\,K would have $k_2$ values
smaller than the BBL-interval (see also \S\ref{sec:Results}).
In the outermost layer of the planet from 1\,bar to 1\,kbar we assume the
temperature profile to be either adiabatic or isothermal. This is supposed to
account for the uncertainty of how the star's radiation influences the outer
envelope of the planet. As the planet is very close to its star, it is likely
that there will be an isothermal layer to some extent \citep{Fortneyetal07}. In
every case, for pressures $P>1$\,kbar the planet is adiabatic.

In addition, we compute a model series that is consistent with the nongray model
atmospheres from \citet{Fortneyetal07} in order to further pin down the outer
boundary condition. Based on their pressure-temperature
profiles of Jupiter-like planets around Sun-like stars \citep[Fig. 3
in][]{Fortneyetal07}, we interpolated a model atmosphere for HAT-P-13b,
assuming that the incoming energy flux remains constant: $L_*/(4\pi a^2) =
\mathrm{const.}$, where $L_*$ is the stellar luminosity and $a$ is the planet's
semi-major axis. Our interpolation yields $T_{1\,\mathrm{bar}} = 2080$\,K as new
outer boundary condition. It further showed that an isothermal layer may reach
down to $P_\mathrm{ad}\sim 100$\,bar. Our interpolated model atmosphere is in
agreement with an atmosphere model specifically calculated for HAT-P-13b, using
the methods of  \citet{Fortneyetal07}.
The thickness of the isothermal layer is a
variable parameter (influenced by the age of the planet). Hence, in our model
series based on the model atmosphere, we vary
$P_\mathrm{ad}$ as a free parameter while keeping $T_{1\,\mathrm{bar}}$ fixed
at 2080\,K.

Our modeling procedure constitutes an improvement compared to the
modeling done by \citet{Batyginetal09} because we do not approximate the core
material by a constant density and we include the effects of different
temperatures and temperature profiles in the atmosphere.
This allows us to draw some general conclusions about the
capability of $k_2$ to constrain interior models of extrasolar giant
planets.

\section{Results} \label{sec:Results}

\subsection{Love number, metallicity, and core mass}

\begin{figure}
 \resizebox{\hsize}{!}{\includegraphics{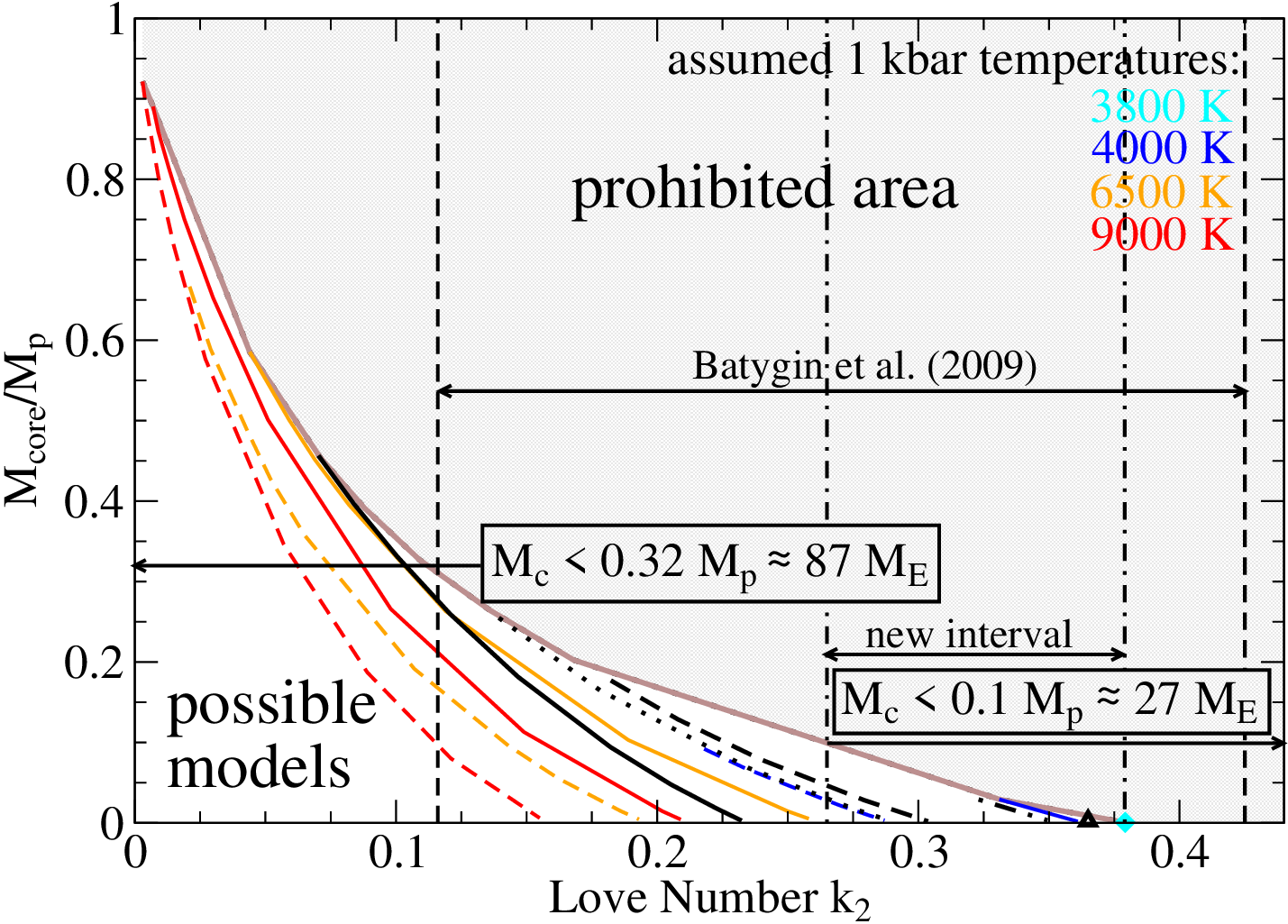}}
 \caption{Love numbers $k_2$ and core masses of two-layer models of HAT-P-13b
 for different 1\,kbar temperatures (color coded). For pressures $<1$\,kbar the
 temperature profile is either adiabatic (colored solid) or isothermal
 (colored dashed). The metallicity increases along a line, reaching its maximum
 value for $M_\mathrm{core}/M_\mathrm{p}=0$. The zero metallicity line for
 fully adiabatic models (brown solid) separates the region of all possible
 planetary models (white) from the prohibited area (light gray).
 Vertical black dashed lines indicate the allowed interval for $k_2$-values
 derived by BBL. Based on that interval we find a maximum possible core mass of
 $0.32\,M_\mathrm{p}\approx87$\,\ME. Our \emph{new $k_2$ interval} is shown
 with vertical black dot-dashed lines. It lowers the uncertainty of the
 planet's core mass down to $0.1\,M_\mathrm{p} \approx 27$\,\ME. The other black
 lines show our models that are based on our model atmosphere with a fully
 adiabatic envelope (solid black) and $P_\mathrm{ad}$ of 5\,bar (dotted),
 10\,bar (dashed black), 50\,bar (dot-dashed black), and 72\,bar (black
 triangle).}
 \label{fig:HAT-P-13b}
\end{figure}

\begin{figure}
 \resizebox{\hsize}{!}{\includegraphics{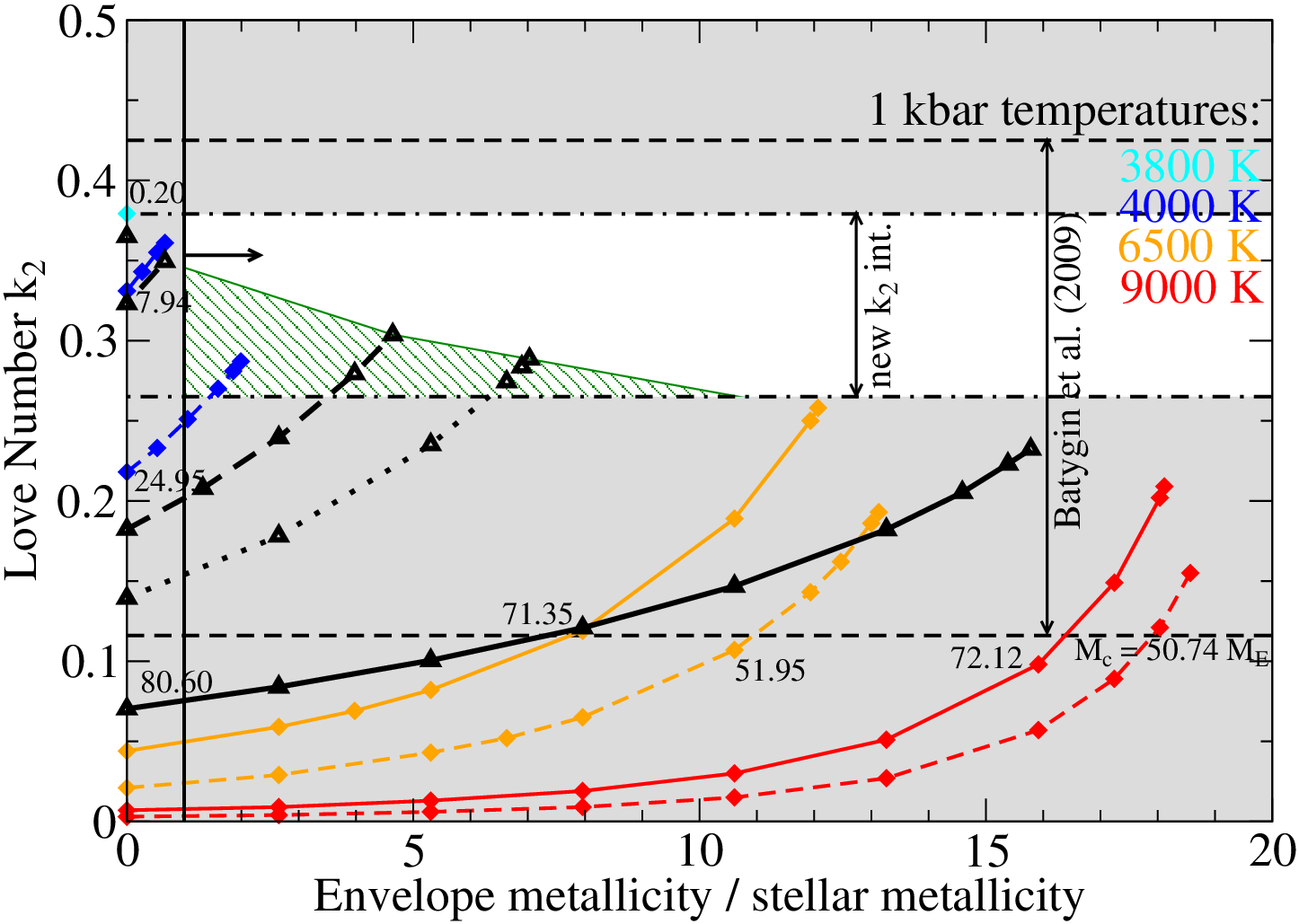}}
 \caption{Love numbers $k_2$ as a function of the envelope metallicity. Shown
are the same models as in Fig.~\ref{fig:HAT-P-13b}. The numbers close to some
models give the core mass in \ME. The shaded areas demonstrate how models can be
ruled out by taking into account the $k_2$ intervals and assuming that the
envelope metallicity is at least the stellar metallicity. Considering the
information from the model atmosphere the favored models reduce to the green
shaded area.}
 \label{fig:HAT-P-13b_k2Z}
\end{figure}

We present in Figs.~\ref{fig:HAT-P-13b} and \ref{fig:HAT-P-13b_k2Z} results for
the calculated core mass, $k_2$, and envelope metallicity. 
The area between an adiabatic and isothermal line of the same 1\,kbar
temperature
accounts for the uncertainty about the outer temperature profile as the
thickness of a potential isothermal layer is not known. Also
shown are the models based on our model atmosphere with an isothermal layer of
2080\,K, reaching down to 1 (fully adiabatic), 5, 10, 50, or
72\,bar, where the core disappears.
In Fig.~\ref{fig:HAT-P-13b}, the line connecting zero metallicity
fully adiabatic models separates the region of all possible planetary models
from the prohibited area which is not accessible for models of HAT-P-13b. 

%metallicity and Love number
A general trend that can be seen in Fig.~\ref{fig:HAT-P-13b}
and~\ref{fig:HAT-P-13b_k2Z} is that the Love
number $k_2$ increases as more metals are put in the envelope. This is the
result of $k_2$ being a measure of the level of central condensation of an
object. Since the total mass of the planet must always be $M_\mathrm{p}$, the
enrichment of the envelope with metals leads to a decrease in core mass. A
smaller core and higher metal content in the envelope means that the planet is
more homogeneous, which is why the Love number grows. The maximum metallicity is
reached when the core of the planet vanishes. This marks the
maximum possible Love number of a planet for a given temperature.

%temperature and Love number
The temperature itself also has a significant influence on the metallicity,
core mass, and consequently Love number of the planet.
Higher temperatures reduce the Love number $k_2$. The higher the envelope
temperature, the lower its density. Low densities in
the envelope require
a more
massive core in order to ensure mass conservation. High envelope temperatures
thus lead to a strong central condensation, reflecting in a small Love number.
For HAT-P-13b we find a minimum 1\,kbar temperature of 3800\,K for fully
adiabatic models. At this low temperature the zero metallicity envelope is
so dense that the core is almost vanished ($M_\mathrm{core}=0.2$\,\ME). Hence,
there is no enrichment of heavy elements possible in the envelope. At this
temperature maximum homogeneity is reached. That translates into a maximum
possible Love number for HAT-P-13b of $k_2=0.379$, well below the upper limit
given by the BBL-interval. At high temperatures, on the other hand, the density
in the envelope is low enough to enable the existence of massive cores. That
creates the possibility to enrich the envelope with metals. For instance, at
$T_\mathrm{1\,kbar}=9000$\,K the core disappears at an envelope metallicity as
high as 70\%. The big cores of the hot models lead to very small Love numbers
and tend to lie outside of the BBL-interval.

We note that models with an isothermal atmosphere generally yield
smaller Love numbers than their adiabatic counterparts, because they are
overall hotter. 

%How models can be constrained or ruled out
As is obvious from Fig.~\ref{fig:HAT-P-13b} and~\ref{fig:HAT-P-13b_k2Z} not all
calculated models are placed in the BBL-interval or even in our narrower
new $k_2$ interval, which means some of the models can be ruled out. The hot
\emph{and} metal poor models cannot fulfill the BBL-interval. Our new $k_2$
interval also rules out the metal rich hot models and only leaves us with the
colder models.

Further constraints can be made by taking into account model atmospheres in
order to get more information about the planet's outer temperature profile.
Our model atmosphere yields a 1\,bar temperature of 2080\,K.
An isothermal layer of $P_\mathrm{ad}>3$\,bar is
required to produce solutions in our new $k_2$ interval.

Another point we can take into account is that planets form out of the same
cloud of gas as their host star. Hence, it is reasonable to assume that the
planet's bulk metallicity is at least the stellar metallicity. Of
course, a
planet
may also be enriched in metals due to capture of planetesimals during formation.
Enhanced bulk metallicities for extrasolar giant planets over their
parent star values were indeed found \citep{MillerFortney11}.
We use this argument to also rule out interior
models that have less than stellar metallicity (see
Fig.~\ref{fig:HAT-P-13b_k2Z}). Here we can approximate the bulk
metallicity with the envelope metallicity as the disputable models have such
small cores that most of the planet's metals are in the envelope. This leaves
us only with very few models
for the interior of HAT-P-13b, where we favor the ones compatible with the
model atmosphere. Combining all the pieces of information, we can
conclude from Fig.~\ref{fig:HAT-P-13b_k2Z} that HAT-P-13b has an isothermal
outer layer with a temperature of 2080\,K that may reach down to 3 --
44\,bar. Models with isothermal layers that extend to higher pressures
have such small cores that their envelope metallicity cannot exceed stellar
metallicity. Consequently, our models with $P_\mathrm{ad} \ge 50$\,bar are
ruled out. The envelope metallicity of the allowed models is
1 -- 11 times the stellar metallicity. Also, the bulk metallicity
cannot exceed this limit as $M_\mathrm{core}<27$\,\ME.

\begin{figure}
 \resizebox{\hsize}{!}{\includegraphics{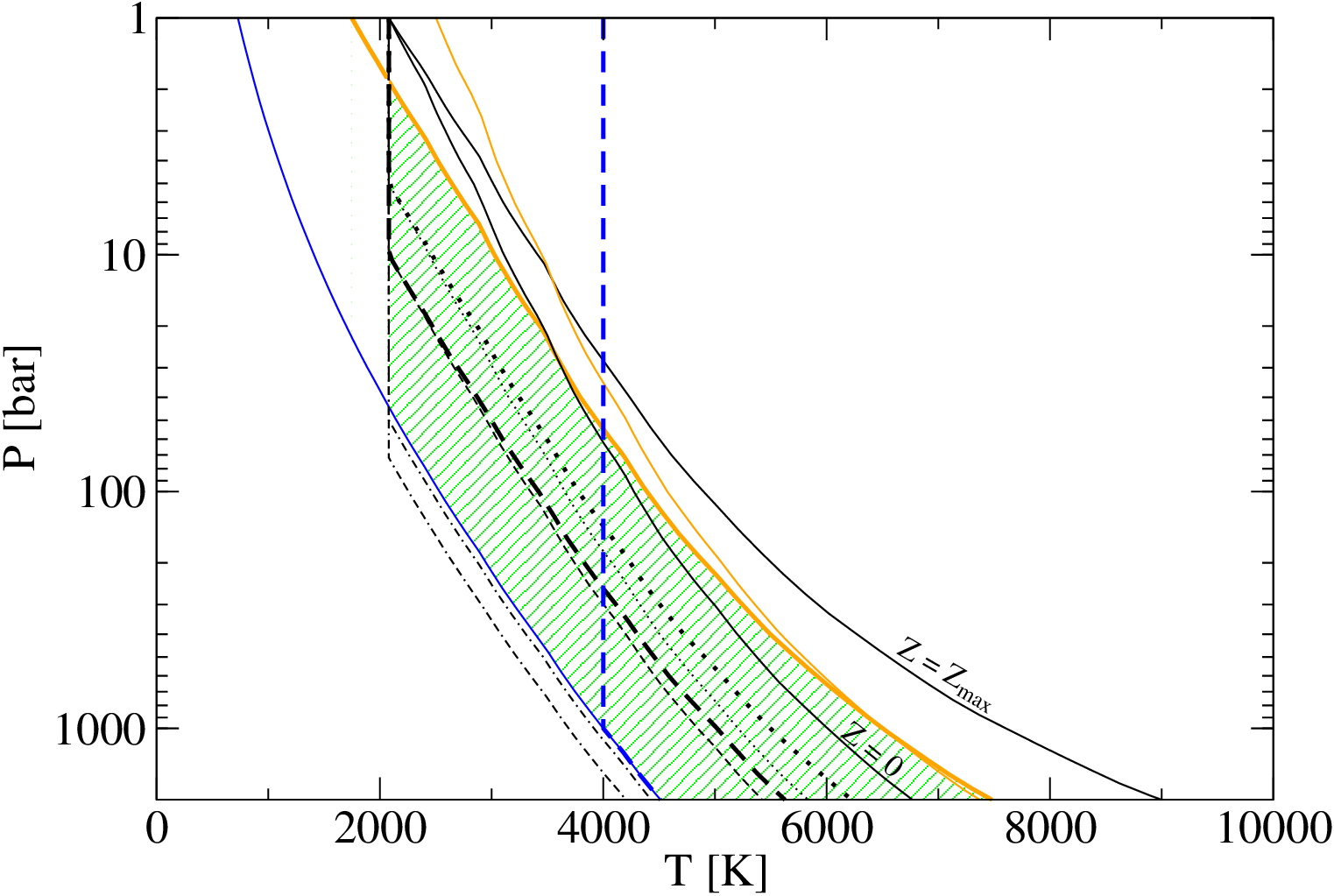}}
 \caption{P--T profiles near the atmosphere of acceptable and unacceptable
models. Colors and line styles are the same as in Figs.~\ref{fig:HAT-P-13b}
and~\ref{fig:HAT-P-13b_k2Z}. For lines with same color and style, models with
$Z=0$ and $Z=Z_\mathrm{max}$ are shown. Thick lines indicate acceptable models;
compare these also to Fig.~\ref{fig:HAT-P-13b_k2Z}. The green shaded area marks the
P--T regime for acceptable models. However, models in this regime are only
acceptable when they also have a large enough amount of metals in their
envelopes.}
 \label{fig:PTallmodels}
\end{figure}

In order to give a better overview of the different boundary conditions
used we plot in Fig.~\ref{fig:PTallmodels} P--T profiles of acceptable and some
unacceptable models. Those with $T_\mathrm{1\,kbar}<4000$\,K and
$T_\mathrm{1\,kbar}>6500$\,K are too cold or too hot, respectively, and hence
not shown in the plot. We find an area that roughly constrains the
P-T-conditions near the atmosphere that produce acceptable models. In addition,
a model also needs to have a high enough envelope metallicity in order to be
acceptable. For example, our structure models compatible with our model
atmosphere with
$T_\mathrm{1\,bar}=2080$\,K and an isothermal layer down to 10\,bar lie
in the
favored area marked by Fig.~\ref{fig:PTallmodels}. However, models of this
kind with $Z=0$ are not acceptable whereas models with a high enough envelope
metallicity are, compare also to Fig.~\ref{fig:HAT-P-13b_k2Z}.

%maximum possible core masses
A very important parameter describing the interior of a planet is its core
mass. This quantity is of interest because it may give a hint on the formation
history of the planet. Figure~\ref{fig:HAT-P-13b} shows that the relationship
between a $k_2$ value and a core mass is non-unique. If, for instance,
$k_2=0.2$ it could either be a very hot $T_\mathrm{1\,kbar}=9000$\,K, metal rich
planet with a very small core or a lower temperature
$T_\mathrm{1\,kbar}=6500$\,K one with a core constituting 10\% of the planet's
total mass. The only information that can be concluded is a \emph{maximum}
possible core mass, given by the line consisting of the adiabatic zero
metallicity models. The maximum possible
core mass of HAT-P-13b is given by the intersection of this line with the lower
$k_2$ boundary, yielding $M_\mathrm{core}<0.32\,M_\mathrm{p}\approx87$\,\ME~for
the BBL-interval. Taking into account our \emph{new $k_2$ interval} lowers the
uncertainty of the planet's core mass down to
$M_\mathrm{core}<0.1\,M_\mathrm{p}\approx27$\,\ME.

In principle, the area under each curve can be filled up with three-layer models
(not shown in Fig.~\ref{fig:HAT-P-13b}). Every redistribution of metals in the
envelope would either result in a stronger central condensation (smaller $k_2$)
or, if central condensation is kept constant, require a transport of material
from the core to the envelope reducing the core mass. As a result of
this degeneracy introduced by a discontinuity in the envelope
\citep[see also][]{Krammetal11} it
is only possible to determine a \emph{maximum} possible core mass.
Further narrowing of the maximum core mass could also be achieved with
a more detailed
measurement of the eccentricity of HAT-P-13b as that would lead to a smaller
$k_2$ interval.

Finally, from all the interior models we presented here, we favor the
ones with
$T_\mathrm{1\,bar}=2080$\,K, an isothermal outer layer of 3 -- 44\,bar, and
an envelope metallicity of $1 < Z/Z_* < 11$ because they (i) are placed in our
\emph{new $k_2$ interval}, (ii) are consistent with the model atmospheres from
\citet{Fortneyetal07}, and (iii) have an envelope and bulk metallicity
that is above the
stellar metallicity. Assuming these parameters are ''correct'' for HAT-P-13b,
the planet would have a core mass of 27\,\ME~or less. For comparison,
Jupiter models from \citet{FortneyNettelmann10} predict
$M_{Z,\mathrm{tot}}=15-38$\,\ME, $M_\mathrm{core}=0-10$\,\ME~and
$Z_\mathrm{p}=4-12\,Z_\odot$.

\subsection{Mass and radius uncertainties} \label{subsec:MRuncertainties}

We now investigate the effect of the measurement uncertainties for the
mass and radius of HAT-P-13b. For that purpose, we varied the mass and radius
within the $1\sigma$ error-bars to obtain different mean densities. As an
example,
we only look at models consistent with our model atmosphere and
$P_\mathrm{ad}=1$ and 50\,bar because all other valid models are bounded
by these and react the same way on mass-radius variations.

\begin{figure}
 \resizebox{\hsize}{!}{\includegraphics{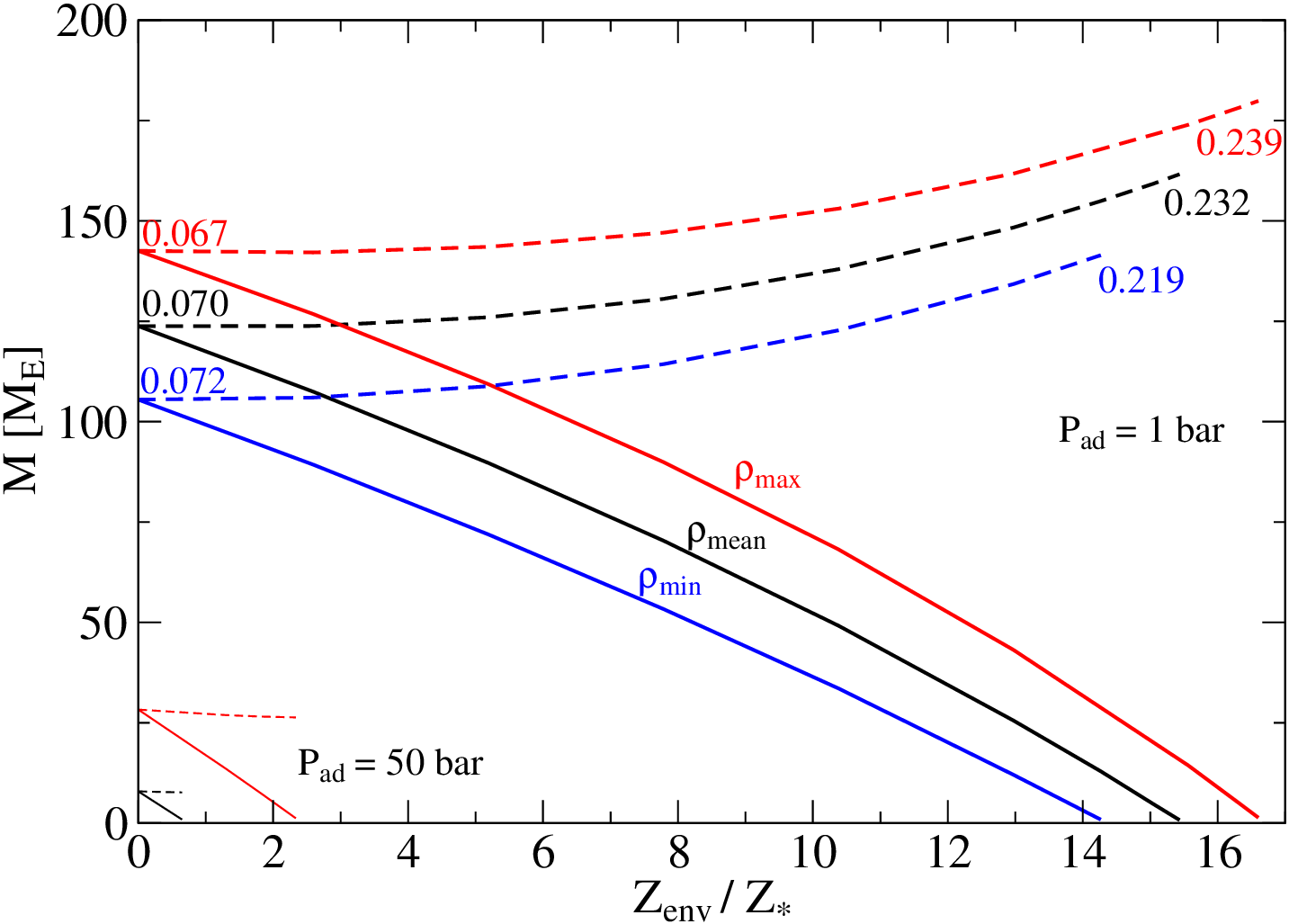}}
 \caption{Core mass (solid lines) and total mass of metals (dashed
lines) as a function of envelope metallicity for different mean densities
according to the measurement uncertainties in mass and radius. The numbers are
the Love numbers $k_2$ for zero and maximum possible metallicity in the
envelope. Shown are only models compatible with our model atmosphere
($T_\mathrm{1\,bar}=2080$\,K) for $P_\mathrm{ad}=1$\,bar and
$P_\mathrm{ad}=50$\,bar.}
 \label{fig:MRvaryZ}
\end{figure}

Figure~\ref{fig:MRvaryZ} shows the effect on core mass and metal
enrichment. The uncertainty in mean density results in an uncertainty of about
$\pm 20$\,\ME~ in core mass and mass of metals. For a given core mass, the
envelope metallicity can vary up to $\pm 3 \,Z_*$.

There is little effect on the Love number $k_2$. In
Fig.~\ref{fig:HAT-P-13b}, the line separating the prohibited area from the
possible models consists of zero-metallicity models. Since for
$Z_\mathrm{env}=0$ the Love number $k_2$ only varies by about $\pm 0.003$, only
a minor shift of this line can be expected so that all the conclusions drawn
for the fiducial $M_\mathrm{p}$ and $R_\mathrm{p}$ values remain valid also
for other mean densities within the error bars.

\subsection{Estimating the tidal $Q$ value}

With $R_\mathrm{p}=1.281$\,\RJ~HAT-P-13b has a large radius given its
mass and age. This implies it possesses an interior energy source, like many
other Hot Jupiter planets. Given its eccentric orbit, this energy source could
have a strong contribution due to tidal heating.

Based on our analysis of $k_2$ and our interior models we found that the
adiabatic transition is between 3 and 44\,bar. We ran atmosphere models to
estimate what values of the intrinsic temperature ($T_\mathrm{int}$) these
would correspond to. At the adiabatic transitions of 3 and 44\,bar we found
$T_\mathrm{int}=750$ and 275\,K, respectively. This also gives us the
intrinsic luminosity $L_\mathrm{int}$. Assuming $L_\mathrm{int}$ is provided
only by tidal heating, we estimated $Q$ with the following relation
\citep{Milleretal09}

\begin{equation}
 L_\mathrm{int}\approx P_\mathrm{t}=\frac{63}{4}\left[(G
M_*)^{3/2}\left(\frac{M_* R_\mathrm{p}^5
e_b^2}{Q}\right)\right]a^{-15/2} \quad.
\end{equation}

This gives us a lower limit on the $Q$ value of HAT-P-13b since tidal
heating might not be the only interior energy source. There could be other
effects contributing to $L_\mathrm{int}$, for instance Ohmic dissipation
\citep{BatyginStevenson10}.

Our estimates of the lower limit on $Q$ for the different
eccentricity measurements are displayed in Tab.~\ref{tab:Qs}. For comparison,
Jupiter's $Q$ value has been calculated to be between $10^5$ and $10^6$
\citep{GoldreichSoter66}. 

\begin{table}
 \caption{Lower limits for the tidal $Q$ value of HAT-P-13b for different
eccentricity measurements and intrinsic temperatures (resulting from the
allowed extension of the isothermal layer in the planet's atmosphere).}
 \label{tab:Qs}
 \centering
 \begin{tabular}{ccc}
  \hline\hline
   $T_\mathrm{int}$ [K] & $e_b=0.021$ & $e_b=0.0133$ \\
     & \citep{Bakosetal09} & \citep{Winnetal10}\\
  \hline
  275 (44\,bar) & 176\,485 & 70\,790 \\
  750 (3\,bar) & 3\,190 & 1\,280\\
  \hline
 \end{tabular}
\end{table}

\section{Discussion} \label{sec:Discussion}

\subsection{Comparison with previous results}
\label{subsec:comparisonwithBatygin}

In addition to an allowed $k_2$-interval, \citet{Batyginetal09} also provided a
set of possible interior models of HAT-P-13b that give $k_2$ values within the
given interval. These models range in core mass from 0 to 120\,\ME. However,
for the given BBL-interval we find a maximum possible core mass that is
significantly smaller ($\approx87$\,\ME). This difference is a result of
different model assumptions. The models of \citet{Batyginetal09} all have cores
with constant densities of 7-12\,g/cm$^3$ (Batygin, \emph{pers. comm.}),
resembling water cores. In contrast, the models presented in this work were
calculated using a compressible rock EOS \citep{HubbardMarley89}, increasing the
densities in the cores up to 12-40\,g/cm$^3$. As a
consequence of the lower and constant core densities in the BBL models, a core
of a specific mass requires a larger radius than a core of the same mass in our
models. This results in a greater Love number than our models with the
same core mass. That is why our models need smaller core masses in order to fall
in the allowed $k_2$-interval. Models of HAT-P-13b in this work with a core mass
of 120\,\ME would have a smaller $k_2$ than the lower $k_2$-boundary of the
BBL-interval. Hence, the different maximum core masses in this work and in
\citet{Batyginetal09} arise from the different core EOS used.
This shows the major influence of the EOS.
In our example
here, the different core EOS cause an uncertainty of about 28\% in the
maximum possible core mass.

\subsection{The eccentricity issue}

As mentioned before (see \S~\ref{subsec:e}) the eccentricity is crucial for
determining an allowed $k_2$-interval. We used the results from BBL to
obtain a \emph{new $k_2$ interval} from another eccentricity measurement
\citep{Winnetal10}. Our new interval, ranging from 0.265 to 0.379
is significantly smaller than the one previously used by BBL ($0.116 < k_2 <
0.425$). This allowed for a more precise estimate of a maximum possible core
mass ($M_\mathrm{core}<27$\,\ME).

It of course has to be kept in mind that the ultimate significance of this new
eccentricity value (and deduced $k_2$) is unclear. Other effects like the
stellar jitter produce a large uncertainty on the eccentricity measurement of
the inner planet, as recently pointed out by \citet{PayneFord11}.

One way to help to constrain the eccentricity would be a detection of
the occultation (secondary eclipse) of the planet (for instance, with \emph{Spitzer}). The
timing of the occultation can be used to place a strong constraint on
$e\cos\omega$, where $e$ is the orbital eccentricity and $\omega$ is the
longitude of periastron \citep{Charbonneauetal05}. Observations to determine
$e\cos\omega$ for HAT-P-13b are currently being made (Harrington \& Hardy,
\emph{pers. comm.}).

\section{Summary \& Conclusions} \label{sec:Summary}

In this work, we presented new interior models of the transiting Hot Jupiter
HAT-P-13b with the aim of showing to what extent interior models of extrasolar
giant planets can be constrained by using the tidal Love number $k_2$ in
addition to the known observables mass and radius. We also varied the envelope
temperature and metallicity 
in order to demonstrate the uncertainties
imposed on the inferred interior models.

%maximum possible core mass
One main result of our work is that based on the Love number $k_2$ one cannot
draw a conclusion on the precise core mass of the planet. Only a
\emph{maximum possible}
core mass can be inferred which is given by adiabatic zero-envelope-metallicity
models. 
Taking into account the allowed $k_2$
interval ($0.116<k_2<0.425$) derived from \citet{Batyginetal09}, we find a
maximum value for HAT-P-13b's core mass of 87\,\ME. Further radial velocity
observations from \citet{Winnetal10} gave a new value for the planet's
eccentricity, allowing us to construct a \emph{new $k_2$ interval} of $0.265 <
k_2 < 0.379$. With this smaller interval we were able to constrain the core
mass of HAT-P-13b further down to 27\,\ME.

%effects of temperature and metallicity
We also showed the influence of the envelope's temperature and metallicity.
The coldest possible model defines a maximum possible value for
$k_2$ as this is the most homogeneous model. For HAT-P-13b we find
$k_{2,\mathrm{max}}=0.379$.
This is well below the
upper $k_2$ limit from \citet{Batyginetal09}.

As our analysis showed, the temperature and metallicity of the planet's envelope
are important parameters whose knowledge would significantly increase the
effectiveness of constraining the planet's interior with the Love number $k_2$.
We therefore calculated a model series based on a theoretical model atmosphere
from \citet{Fortneyetal07}.
Models of this kind have $T_\mathrm{1\,bar}=2080$\,K and fall in our \emph{new
$k_2$ interval} if they have an outer isothermal layer with $3\,\mathrm{bar} <
P_\mathrm{ad} < 44\,\mathrm{bar}$. With this result we also estimated
lower limits of the tidal $Q$ value of HAT-P-13b.

Based on our analysis and the applicability of the tidal fixed point
theory we can rule out the majority of the calculated models.
Assuming our new $k_2$-interval, we
note that hot models with $T_{1\,\mathrm{kbar}}>6500$\,K can be ruled out
(compare Fig.~\ref{fig:HAT-P-13b_k2Z}). On the
other hand, the very cold models with $T_\mathrm{1\,kbar}<4000\,$K are
unlikely because they have a metallicity that is smaller than stellar.
We favor models with
$T_\mathrm{1\,bar}=2080$\,K, an isothermal outer layer of 3 -- 44\,bar, and
an envelope metallicity of $1 < Z/Z_* < 11$ because they (i) are placed in our
\emph{new $k_2$ interval}, (ii) are consistent with the model atmospheres from
\citet{Fortneyetal07}, and (iii) have an envelope metallicity that is above the
stellar metallicity. Assuming these conditions really apply for
HAT-P-13b the planet would have a core mass of
27\,\ME~or less. Given the calculated tidal $Q$, if HAT-P-13b turns out
to have a small core and an envelope enrichment similar to Jupiter, it could
represent a Jupiter-like extrasolar planet.

By comparing with the previous results from
\citet{Batyginetal09} we have seen that the core EOS affects the core mass
and Love number, resulting into a difference of about 28\% in the prediction of
a maximum possible core mass of HAT-P-13b.

%discussion about e
As pointed out in \S\ref{subsec:e}, the eccentricity of the inner planet is a
crucial parameter because it determines the Love number $k_2$. The value of the
eccentricity is still uncertain. We analyzed the consequences of an
$e_b = 0.021\pm0.009$ \citep{Bakosetal09} and $e_b = 0.0133\pm0.0041$
\citep{Winnetal10}. 

%motivate further observations
Finally, despite the uncertainty of the inner planet's eccentricity it is still
unclear whether the tidal fixed point theory from \citet{Mardling07} can really
be applied to HAT-P-13b. Further observations are necessary to prove the
co-planarity and apsidal alignment configuration of the system. Also, TTV
measurements could provide clues about the existence of other long-period
companions in the system.

\begin{acknowledgements}
We thank Konstantin Batygin, David Stevenson, Josh Winn and Stefanie Raetz for
helpful discussions. UK, RR, and RN acknowledge support from the DFG SPP 1385
''The first ten million years of the Solar System'', NN from DFG
RE882/11-1.
\end{acknowledgements}

\appendix

\bibliographystyle{aa} % style aa.bst
\bibliography{references} % your references Yourfile.bib

\section{Applicability of the tidal fixed point theory}
\label{sec:appfixedpoint}

% if not co-planar --> limit cycle (Mardling)
The analysis in this work is based on the assumption that the tidal
fixed point theory
of \citet{Mardling07} can be applied in order to extract a $k_2$ value from the
inner planet's eccentricity. For that purpose the
planetary system has to fulfill the requirements mentioned in
\S\ref{subsec:HATP13b-obsconst}. Yet unproven and crucial to the evolution
into the tidal fixed point is the co-planarity of the HAT-P-13 system. In fact,
if the orbits of planet b and c are mutually inclined, the system will not
evolve to the tidal fixed point. Instead it will relax to a \emph{limit cycle}
in $e_b$ - $\eta$ parameter space where the eccentricity of the inner planet
$e_b$ and the angle between the periapse lines $\eta$ oscillate
\citep{Mardling10}. In that case the Love number $k_2$ of the inner planet can
not be determined unambiguously. In particular, \citet{Mardling10} shows that
the analysis of \citet{Batyginetal09} and hence the inferred $k_2$-interval is
only valid if the mutual inclination of planets b and c is less than about
10$^\circ$.

% how to prove co-planarity, transit of c and psi (Winn)
A direct estimate of the mutual inclination $\Delta i$ would be possible
by observing planet c in transit and measuring the Rossiter-McLaughlin effect.
A transit of HAT-P-13c was predicted to occur on April 28, 2010
\citep{Winnetal10}. However, such a transit could not be observed. The
multi-site campaign by \citet{Szaboetal10} revealed that a transit of HAT-P-13c
can be excluded with a significance level from 65\% to 72\%. Yet, a close to
co-planar configuration cannot be ruled out. The transit probability of
HAT-P-13c is at most 8.5\% for a mutual inclination of the two planets of
8$^\circ$ and decreases to zero for inclinations $\lesssim 4^\circ$
\footnote{For comparison, the solar system planets are inclined within
3.4$^\circ$ of the Earth's orbit (except for Mercury)} \citep{BeattySeager10}.
Another reason why the campaign of \citet{Szaboetal10} could not detect a
transit may be that they only observed for 5 days. The orbital parameters
(eccentricity and period) of the outer planet are not well constrained due to
the effect of the unknown stellar jitter, which translates into large
uncertainties about the predicted transit times of planet c, making it
necessary to observe over a wide range of nights--- $\sim3$ weeks
\citep{PayneFord11}.

Support for the hypothesis of co-planarity of the HAT-P-13 system comes from the
study of the Rossiter-McLaughlin effect to determine the stellar obliquity
$\Psi_{*,b}$. \citet{Winnetal10} note that there is an indirect connection
between $\Psi_{*,b}$ and $\Delta i$: a large $\Delta i$ would lead to nodal
precession of b's orbit around c's orbital axis, causing periodic variations in
$\Psi_{*,b}$. As a consequence, it would be unlikely to observe a low value
of $\Psi_{*,b}$ unless $\Delta i$ is also small. At any rate, only
the sky-projected
angle $\lambda$ can be measured (not the true obliquity $\Psi_{*,b}$) which
makes it still impossible to draw firm conclusions about $\Delta i$.
\citet{Winnetal10} find a small $\lambda=1.9\pm8.6\,{\rm deg}$, suggesting that
planet b's orbital angular momentum vector is well-aligned with the stellar
spin vector. This also increases the likelyhood that the orbits of planets b
and c are co-planar.

% effect of possible third planet + TTV
\citet{PayneFord11} also suggest a method to determine the mutual
inclination between
planets b and c with the help of transit timing variations (TTV). In
particular, they show that systems with $\Delta i \sim 0^\circ$ and $\Delta i
\sim 45^\circ$ result in significantly different TTV profiles which can
easily be distinguished. In fact, HAT-P-13b has been subject to a TTV campaign.
\citet{Paletal11} detected TTV in the order of 0.01 days which 'suddenly'
occured during their observations in December 2010 and January 2011 while in
the first three years of observations of HAT-P-13b there seemed to be a lack of
TTV. If it is a periodic process, it should have a period of at least
$\sim3$ years, making it unlikely to be caused by HAT-P-13c. Hence,
\citet{Paletal11} conclude that the measured TTV may be the result of
perturbations of
another long-period companion. Interestingly, \citet{Winnetal10} also point to
the possibility of an additional body d in the system as their models of two
Keplerian orbits give an unacceptable fit to the RV data while a model assuming
a third companion with a longer orbital period is successful. Furthermore,
thinking about the origin of the HAT-P-13 system, \citet{Mardling10} suggests
that the system originally contained a third planet which was later scattered
out of the system.

Given all these issues, to our current knowledge it seems impossible to say
whether HAT-P-13b is really subject to the tidal evolution into the tidal fixed
point like envisioned by \citet{Mardling07}. Further observations
are necessary to pin down the
mutual inclination of the planets and to prove or disprove the
existence of another companion in the system.

\end{document}